\documentstyle[12pt]{article}
\newcommand{\be}{\begin{equation}}
\newcommand{\ee}{\end{equation}}
\newcommand{\bea}{\begin{eqnarray}}
\newcommand{\eea}{\end{eqnarray}}

\begin{document}
%%%%%%%%%%%%%Title page%%%%%%%%%%%%%%%%%%%

\begin{center}
\begin{large}
{\bf   Reissner-Nordstrom Near Extremality \\}
{\bf from a\\}
{\bf  Jackiw-Teitelboim Perspective \\}
\end{large}  
\end{center}
\vspace*{0.50cm}
\begin{center}
{\sl by\\}
\vspace*{1.00cm}
{\bf A.J.M. Medved\\}
\vspace*{1.00cm}
{\sl
Department of Physics and Theoretical Physics Institute\\
University of Alberta\\
Edmonton, Canada T6G-2J1\\
{[e-mail: amedved@phys.ualberta.ca]}}\\
\end{center}
\bigskip\noindent
\begin{center}
\begin{large}
{\bf
ABSTRACT
}
\end{large}
\end{center}
\vspace*{0.50cm}
\par
\noindent

In this paper,
we investigate the near-extremal thermodynamics of the Reissner-Nordstrom
(RN) black hole. Our methodology is based on a duality that exists
between the near-horizon geometry of the near-extremal RN sector
and Jackiw-Teitelboim (JT) theory. First, the described correspondence
is reviewed at the classical level. Next, we consider first-order
perturbations in the dual JT geometry by incorporating a quantum scalar
field into the formalism. The novelty of our approach is that the
matter field is endowed with a 4-dimensional pedigree. We  ultimately
find that back-reaction effects prohibit the JT black hole from
losing all of its mass.  This outcome directly implies  
that an RN black hole can not  reach extremality and, moreover, can
 not even come
arbitrarily close to an extremal state.

%PACS 04.70.Dy
\newpage

\section{Introduction}
%\medskip
\par

 A complete  theory of quantum gravity currently remains
a most formidable obstacle. Yet,  much has been ascertained
by considering gravitational effects in a  semi-classical regime.
At the forefront of such research has been the remarkable
analogy between black hole mechanics and thermodynamic systems.
Certainly,  black hole analogies to the zeroth, first and second laws 
of thermodynamics
have been well established  with  the appropriate  identifications: 
 the 
surface area of the black hole  horizon with entropy \cite{bek,haw} and 
the surface gravity at the horizon  with temperature \cite{haw2}. However,
a third law of black hole thermodynamics, although technically formulated 
\cite{isr},   is still in need of much  clarification before
the thermodynamic  analogy can be considered complete \cite{wal}.  
\par
It is worth reviewing the  third law as it applies to
``conventional'' systems. There are actually two formulations
of this law; both of which can be attributed to Nernst \cite{thermo}.
Firstly, there is a formulation stating
that the entropy of a system approaches a constant
(which must be independent of all macroscopic parameters of the system) as the 
temperature
approaches absolute zero. Typically, this constant is taken to be zero. 
Secondly,  there is a formulation stating that it is impossible
to  reach absolute zero by a finite number of reversible
processes.
For ordinary thermodynamic systems, it is easy to show that
these are equivalent statements. However, this is not
the case for black holes as we shall soon see.
\par
For sake of illustration, let us consider a Reissner-Nordstrom
black hole (i.e., charged, spherically symmetric black hole in 4-dimensional
spacetime) with mass $M$ and charge $Q$.
 Examining the so-called extremal limit of  $M^2\rightarrow Q^2$,
we know that the surface gravity (i.e., temperature) goes
as $\kappa \propto \sqrt{M^2-Q^2}$, while the
surface area (i.e., entropy) is given  by $A = 4\pi(M+\sqrt{M^2-Q^2})^2$
(with all fundamental constants set to unity). 
It is immediately clear that the Hawking
temperature vanishes in the extremal limit, so one would expect
that the corresponding entropy would either vanish or approach a constant
parameter. Alas, this is not the case, as the extremal value
of entropy clearly depends on a macroscopic parameter; namely,
the  mass $M$ of the black hole. 
\par
Given the ambiguous nature of the extremal-limiting case, 
one is forced to choose  between two possible  revisions of
the third law as applied to black hole thermodynamics. Either
that  the ``enthropic'' formulation  of the third law
holds in a weakened form (i.e., the zero-temperature limit
does not coincide with a vanishing entropy)
 or there must be  a
discontinuity occurring  near the zero-temperature limit
 (i.e., zero temperature may yield a state of vanishing entropy,
but such a state can not be viewed as
limiting case of a finite-temperature system).   This dilemma has
resulted in two very different schools of thought regarding
 the interpretation of extremal  black holes.
\par
On one hand,  an extremal black hole is often interpreted
as being a well-defined (zero-temperature) limit of its non-extremal 
counterpart. The most compelling argument on behalf of this
viewpoint has come from   string-theory calculations of black hole
entropy. More specifically, Strominger and others \cite{str,hor}
have considered certain classes of weakly coupled
string theory for which massive string states can be represented
by extremal black holes. In these works, a statistical
procedure has been used to generate the Bekenstein-Hawking area
law (i.e., entropy is equal to one quarter of the surface area), precisely.
If such a result is valid, then clearly the extremal limit must
be well-defined with entropy being  a non-vanishing, mass-dependent quantity
in this limit.
\par
On the other hand, it has been conjectured that extremal black holes
and non-extremal black holes are qualitatively distinct entities,
with no chance of one continuously deforming into the other.
In this picture, extremal black holes are indeed assigned zero
entropy, assuming that they can exist at all.
Support for this  point of view has come from various sources.
Here, we take  note of studies that closely
 examined  the following issues:  topological differences
between  non-extremal and extremal geometries \cite{hhr,tei,kr,gk},
extremal breakdowns occurring at perturbative order \cite{bur1,aht,me1},
 infinitley long time periods required to achieve extremality
 \cite{woo1},
non-thermal behavior of an ``incipient'' extremal black hole \cite{lrs}
and violations in the second law for finite-entropy extremal
black holes \cite{hod}.
\par
Suffice it to say, given these two  conflicting viewpoints, there is 
still no consensus  on
the subject of extremal thermodynamics.
\par
Let us now consider a recent paper, of relevance to this subject, by Fabbri,
Navarro and Navarro-Salas \cite{fnn}. These authors  demonstrated
a direct correspondence  between
the near-horizon behavior of 
near-extremal Reissner-Nordstrom (RN) black holes 
and a static geometry in
AdS$ _{2}$ (i.e.,  2-dimensional anti-de Sitter 
spacetime).\footnote{This dual 
behavior  originates, quite naturally, by way of 
the AdS$ _{2}$/CFT$ _{1}$ correspondence  
\cite{x1,x2,x3,x4}.}
What is particularly interesting 
about this correspondence
is that it implies  a duality between near-extremal RN black holes
and near-massless  black holes in Jackiw-Teitelboim (JT) theory \cite{jt}.
Such a duality has intriguing ramifications regarding the third law
of thermodynamics for the following reason.  Massless
JT black holes have both a vanishing temperature and a vanishing entropy.
That is, with regard to JT theory, the third law of thermodynamics
holds naturally in both of Nernst's formulations. Given the
eternal vagueness of what constitutes the ``true physical
picture'' of any given system, this duality  should be taken seriously
as supporting the viewpoint of a  well-defined extremal limit.
\par
Fabbri et al. \cite{fnn} continued on in their paper
with  an analysis of back-reaction effects due to the presence
of minimally coupled scalar fields. On this perturbative level, their 
rigorous analysis
supported a well-defined extremal limit. However, here, we have to 
 take issue with one point. The  matter fields were assumed to be
minimally coupled with regard to the JT black hole theory.
Since the JT action, in this context, has its origins in a 4-dimensional
theory, we feel that any matter fields   should be subjected to
the same criteria. That is, it can be argued that the matter
fields should be, for instance, minimally coupled in the
original 4-dimensional theory. In this case, the form of  coupling
in the dual 2-dimensional theory can only be ascertained
after implementing the same dimensional-reduction ansatz
as  applied to the classical  action.
An investigation into this very matter is the topic of
the current paper.
\par
The content of this paper is organized as follows.
In Section 2, we review the correspondence that was
found between near-extremal RN black holes and
near-massless JT black holes \cite{fnn}.  Then, in Section 3,
we consider quantum back-reaction effects to 
first-perturbative order.  (The philosophy that underlies  our approach 
is based on a study  by Anderson, Hiscock and Taylor \cite{aht}.)
In particular,  the quantum-corrected form 
of the surface gravity (or, equivalently, temperature) is calculated. 
After which, we
determine if  a massless  black hole  complies with 
a non-negative surface gravity; a  necessary condition for 
any black hole thermodynamic system to have physical meaning.
With an eye to completeness, this analysis is carried out
for both a matter field that is minimally coupled
in  the effective 2-dimensional theory and  in the
original 4-dimensional theory.
Note that, for the  sake of brevity, much of the 
technical details of this section
have been  left to  previously published papers
 by  Kunstatter and this author \cite{me1,me2}.  
Finally, Section 4 closes with a summary and discussion of the 
results.

\section{Near-Extremal Reissner-Nordstrom Black Holes}
\par

Fabbri, Navarro and Navarro-Salas \cite{fnn} have recently  demonstrated 
an explicit  correspondence between near-extremal Reissner-Nordstrom (RN)
 black holes
and near-massless black holes of 2-dimensional AdS gravity.
In this section,
we review this duality, as it is essential to
the later analysis of this paper.
\par
Let us begin with the 4-dimensional Einstein-Maxwell action:\footnote{For
the duration of the paper, we work in units such that the speed
of light and Boltzmann's constant are set equal to unity.}
\be
I^{(4)}={1\over 16\pi G}\int d^4x\sqrt{-g^{(4)}}\left[
R^{(4)}-F^{AB}F_{AB}\right],
\label{1}
\ee
where $G$ is the 4-dimensional Newton constant and 
$F_{AB}$ is the Abelian field-strength tensor ($A,B=0,1,2,3$).
\par
The unique static and spherically symmetric solution 
of this action can be described by the well-known RN metric:
\be
ds^2=-\left(1-{2GM\over r}+{GQ^2\over r^2}\right)dt^2
+ \left(1-{2GM\over r}+{GQ^2\over r^2}\right)^{-1}dr^2 +r^2d\Omega^2,
\label{2}
\ee
where $M$ and $Q$ represent the conserved quantities of black hole mass
and black hole charge (respectively).
If $M^2G^2>GQ^2$, this is the solution for a charged, non-extremal black hole
with two distinct horizons. These are given by:
\be
r_{\pm}=GM\pm \sqrt{G^2M^2-G Q^2}.
\label{3}
\ee
\par
The thermodynamic properties of non-extremal, highly symmetric
black holes are  well established \cite{bek,haw,haw2}. In the RN
case, the associated entropy and temperature are respectively found to be:
\be
S_{BH}={A_{+}\over 4\hbar G}={\pi r_{+}^2\over \hbar G},
\label{4}
\ee
\be
T_{H}={\hbar\kappa_+\over 2\pi}=\hbar {r_{+}-r_{-}\over 4 \pi r_{+}^2},
\label{5}
\ee
where $A_+$ is the surface area and $\kappa_+$ is
the surface gravity with respect to the outermost horizon.
\par
For the case of extremal black holes (i.e., when $G^2M^2=GQ^2$ or
$r_-=r_+$),  the
associated thermodynamic properties remain an open issue (as discussed
in  Section 1). However, we can still safely consider a ``near-extremal
regime'' by setting $\Delta M =M- M_{o}$, where
$M_{o}^2=Q^2/G$ (with the charge $Q$
assumed to be a fixed quantity). Then to leading order in $\sqrt{\Delta M}$: 
\be
r_{+}=GM_o +G\sqrt{2M_o \Delta M},
\label{6}
\ee
\be
\Delta S_{BH}\equiv S_{BH}(M,Q)-S_{BH}(M_o)={4\pi G M_o \over \hbar} \sqrt{
{M_o\Delta M\over 2}}
\label{7}
\ee
\be
\Delta T_H\equiv T_{H}(M,Q)-T_{H}(M_o) = {\hbar\over 2\pi G M_o^2}
\sqrt{2 M_o\Delta M},
\label{8}
\ee
where $S_BH(M_o)=\pi GM_o^2/\hbar$ and $T_H(M_o)=0$.
\par
It is well known that imposing spherical
symmetry on  4-dimensional Einstein-Maxwell gravity  leads to
a 2-dimensional effective theory.\footnote{For futher discussion
on the  various aspects of 2-dimensional gravity, one should consult
the series of papers in Ref.\cite{strb} (and references therein).}
 After this procedure, the dimensionally
reduced action can  be expressed as follows (see, for example, 
refs.\cite{lk1,me3}):
\be
I=\int d^2x \sqrt{-g}\left[{\phi^2\over 4l^2}R+ {1\over 2 l^2}(\nabla \phi)^2
+{1\over 2l^2}-{Q^2\over 2 \phi^2}\right ],
\label{9}
\ee
where the ``dilaton'' $\phi$ is identifiable with the radius of the 
symmetric two-sphere, $l^2=G$,
 the conserved charge is (still) given by $Q$, and all 
geometric quantities have been  defined with respect to the resultant
 1+1-dimensional
manifold.
\par 
It is convenient to eliminate the kinetic term in Eq.(\ref{9})
by way  of the following field reparametrization \cite{lk2}:
\be
{\overline \phi}={\phi^2 \over 4 l^2},
\label{9A}
\ee
\be
{\overline g}_{\mu\nu}= \sqrt{{\overline \phi}} g_{\mu\nu}.
\label{10}
\ee
The reparametrized action then takes on the following form:
\be
I=\int d^2x \sqrt{-{\overline g}}\left[{\overline \phi} R({\overline g})+ 
 {1\over l^2} {\overline V}_Q({\overline \phi})\right ],
\label{11}
\ee
where:
\be
{\overline V}_Q({\overline \phi})= {1\over 2 \sqrt{{\overline \phi}}
}\left [1-{Q^2\over 4 {\overline \phi}}\right].
\label{12}
\ee
\par
It  is pertinent to this analysis  that the extremal
configuration (i.e., the extremal RN limit assuming its existence) 
can be recovered
when the ``potential'' ${\overline V}_{Q}({\overline\phi})$
vanishes. That is, when:
\be
{\overline \phi}={\overline \phi}_o\equiv {Q^2\over 4}.
\label{13}
\ee
With this in mind,
let us now define ${\tilde \phi}={\overline \phi}-{\overline \phi}_o$
and expand the action (\ref{11}) about the extremal
configuration. To first order in ${\tilde \phi}$, the following is obtained:
\be
I=\int d^2x \sqrt{-{\overline g}}\left[{\tilde \phi} R({\overline g})+ 
 {1\over l^2} {\tilde V}_Q({\tilde \phi})\right ],
\label{14}
\ee
where:
\be
{\tilde V}_Q({\tilde\phi})=\left.
{d {\overline V}_Q \over d{\overline\phi}}\right|_{{\overline\phi}_o}
{\tilde \phi}={4\over |Q|^3}{\tilde\phi}.
\label{15}
\ee
\par
Henceforth, we drop the tildes and bars;  thus considering
the following action:
\be
I=\int d^2x \sqrt{- g}\phi\left[ R( g)+ 
 2{\lambda\over l^2} \right ],
\label{16}
\ee
where $\lambda=2/|Q|^3$.
This is simply the action for 2-dimensional  AdS gravity
for which black hole solutions are known and commonly referred to
as Jackiw-Teitelboim (JT) black holes \cite{jt}.
\par
It can be readily shown that, for a static gauge,
the general  solution of the JT action (\ref{16})
is expressible as follows:
\be
ds^2= -(\lambda {x^2\over l^2}-ml)dt^2+(\lambda {x^2\over l^2}-ml)^{-1}
dx^2,
\label{17}
\ee
\be
\phi={x\over l},
\label{18}
\ee
where $m$ represents the conserved mass of the JT black hole.
Moreover, with straightforward application  of Ref.\cite{lk3}
(applicable to  a generic 2-dimensional dilaton  theory),
we are able to identify the following thermodynamic properties:
\be
S_{JT}={4\pi\over \hbar} \phi_{+},
\label{19}
\ee
\be
T_{JT}={\hbar\lambda\over 2\pi l}\phi_{+},
\label{20}
\ee
where $\phi_{+} = x_{+} /  l =\sqrt{lm / \lambda}$
is the horizon value of the dilaton field.
\par
Recalling that $\lambda=2/|Q|^3$, $l^2= G$ and $Q^2=GM_o^2$,
and also identifying $m$ with $\Delta M$, we 
can easily show the following:
\be
S_{JT}= {4\pi G M_o\over\hbar}\sqrt{{M_o\Delta M\over 2}},
\label{21}
\ee
\be
T_{JT}= {\hbar\over 2\pi G M_o^2}\sqrt{2 M_o \Delta M}.
\label{22}
\ee
A direct comparison of these outcomes with Eqs.(\ref{7},\ref{8}) yields a
 couple of intriguing identifications:
\be
S_{JT}=\Delta S_{BH},
\label{23}
\ee
\be
T_{JT}=\Delta T_{H}.
\label{24}
\ee
That is, the thermodynamic properties of a near-extremal black 
hole coincide with JT thermodynamics. 
\par
The massless ($m\rightarrow 0$) limit  
in the JT sector appears to be a well-defined
limiting procedure. Moreover, in this massless limit,  
$S_{JT}\rightarrow 0$ as $T_{JT}\rightarrow
0$; so  the enthropic formulation  of the third law of thermodynamics
has  been  explicitly realized. With these observations, as
well as the observed duality, 
it is tempting to conclude  that  extremal RN black holes
can be interpreted as a limiting case of  non-extremal
 solutions. However, such a conclusion is premature
until back-reaction effects have  properly been  accounted for.

\section{Perturbed Jackiw-Teitelboim Black Holes}
\subsection{General Setup}
\par
Generally speaking,
the emission of Hawking radiation is expected to have repercussions
on the underlying black hole  geometry.  This effect should be of 
particular importance near the extremal (or, for JT theory, massless)
limit, where even the smallest changes in black hole mass
can significantly deform the background geometry. 
To investigate the implications of such  back-reaction effects,
we will suitably adapt an approach suggested by Anderson et al. \cite{aht}.
\par
Let us begin here by considering a JT black hole in a state of thermal
equilibrium with a quantized matter field. This  equilibrium state
ensures that the perturbed geometry continues to be static and,
hence, can be generically expressed in the following manner:
\be
ds^2= -e^{2\omega(x)}\left[\lambda {x^2\over l^2}-ml-l\mu(x)\right]dt^2
+\left[\lambda {x^2\over l^2}-ml-l\mu(x)\right]^{-1}
dx^2.
\label{25}
\ee
With this ansatz, the quantum corrections are expressed in terms of a 
``mass correction'' $\mu(x)$ and another function $\omega(x)$;
both of which are  required to vanish in the classical 
($\hbar\rightarrow 0$) limit. For future convenience, we will introduce
a perturbative parameter $\epsilon\sim \hbar << 1$ and 
write:
\be
\mu(x)=\epsilon\mu_{1}(x)+{\cal O}(\epsilon^2),
\label{25A}
\ee
\be
e^{2\omega(x)}=1+2\epsilon\omega_1(x)+{\cal O}(\epsilon^2).
\label{25B}
\ee
\par
Let us next consider the semi-classical Einstein equation:
\be
G^{\mu}_{\nu}=<T^{\mu}_{\nu}>.
\label{26}
\ee
Using the quantum-corrected metric (\ref{25}) to describe
the Einstein tensor ($G^{\mu}_{\nu}$), one finds
the following to first order in $\epsilon$ (see, for example, Ref.\cite{fro}):
\be
\epsilon \mu_1^{\prime}= -<T^t_t>,
\label{27}
\ee
\be
\epsilon \omega_1^{\prime}= {l^3\over 2(\lambda x^2-l^3 m)}\left[
<T^x_x>-<T^t_t>\right],
\label{28}
\ee
where primes indicate differentiation with respect to $x$.
\par
To check the consistency of the perturbed solution when near extremality,
it is useful to consider the quantum-corrected surface gravity.
The premise being that black hole thermodynamics
can only have physical meaning when the surface gravity
(or, equivalently, the temperature) maintains a non-negative
value.  Using standard calculational procedures \cite{wal2}, we have:
\be
\kappa = {e^{-\omega}\over 2}\left.\left[\lambda {x^2\over l^2}
-lm-l\mu\right]^{\prime}\right|_{x=x_{+} +  q},
\label{29}
\ee
where $x_+=\sqrt{l^3 m/ \lambda}$ is the classical horizon and 
$q$ is its quantum deformation.
Up to first order in $\epsilon$, this expression can be written:
\be
\kappa =\left[1-\epsilon\omega_1(x_+)\right]{\lambda\over l^2}x_+
+\epsilon {\lambda\over l^2} q_{1}-\epsilon{l\over 2}\mu^{\prime}_1
(x_+),
\label{30}
\ee
where $q=\epsilon q_{1} +{\cal O}(\epsilon^2)$.
\par
In principle, $\mu_1$ and $\omega_1$ can be directly obtained from
Eqs.(\ref{27},\ref{28}), while the one-loop horizon
shift can be expressed as:\footnote{The relation follows
from a Taylor expansion of the defining relation:
$l^{-2} \lambda (x_+ + q)^2-lm-l\mu =0$.}
\be
q_1={l^3\over 2\lambda x_+}\mu_1(x_+).
\label{31}
\ee
\subsection{Minimal Coupling in 2-D}
\par
If we are to proceed with an explicit calculation, it
is necessary to make some assumptions regarding the
nature of the quantized matter field.
Let us begin with the simple choice of a massless
scalar field ($f$) that is minimally coupled in
the 2-dimensional theory. The revised (total) action thus becomes:
\be
I_{TOT}=I_{JT}-{\hbar\over 2}\int d^2 x \sqrt{-g}(\nabla f)^2,
\label{302}
\ee
where $I_{JT}$ is the action of Eq.(\ref{16}). 
After integrating out the matter field and then taking the vacuum 
limit, one is known to obtain a quantum effective action of
the following form \cite{pol}:
\be
I_{TOT}=I_{JT}-{\hbar\over 96\pi}\int d^2 x R{1\over\Box}R.
\label{303}
\ee
\par
At this point, we defer the details of the stress-tensor calculation
(and related quantities)
to Ref.\cite{me2}.  In this prior study,  first-order quantum corrections to
the JT model 
were explicitly calculated  for the case of minimally coupled scalar
fields.  The following results, which are of pertinence to
the current study, were found:\footnote{There is a minor
modification to the prior results due to the extra
factor of $\lambda$ in the cosmological-constant term.
This additional consideration only trivially affects
the calculations.}  
\be
\epsilon \mu^{\prime}_1 (x_+)= {\hbar\lambda\over 24 \pi l^2},
\label{306}
\ee
\be
\epsilon\omega_1(x_+)=0 + {\cal O}\left({1\over L}\right),
\label{307}
\ee
\be
\epsilon q_1={\hbar l \over 48\pi}.
\label{308}\ee
Here,  $L$ is the extent of the system's outer 
boundary,
which (for our purposes) can be taken to 
infinity.\footnote{Since
we are considering a black hole in thermal equilibrium, it
is necessary to enclose the system in a ``box'' 
\cite{york}.}    \footnote{One might 
anticipate a similar box contribution arising
in the calculation of $\mu_1(x_+)$ and (hence) $q_1$.
However, in this case, we have assumed the constant of integration
to be absorbed into the ``renormalized'' zeroth-order
mass ($m$).}
Substituting these results into Eq.(\ref{30}) for the first-order
 surface gravity, we have:
\be
\kappa ={\lambda\over l^2}x_+ = \sqrt{{\lambda m\over l}}.
\label{309}
\ee
\par
To review, we have considered the case of a minimally coupled 
 scalar field; that is, minimally coupled with regard
to the  2-dimensional JT theory. We find that
 the surface gravity  does indeed remain non-negative,
even in the massless limit,
when first-order perturbative effects are considered. 
However, as priorly discussed, this is not necessarily
an appropriate form for the  coupling. In Section 1, we have argued
that the matter fields should have some sort of  4-dimensional pedigree.
One  simple, sensible choice is the  condition of a minimally
coupled scalar in the 4-dimensional model.  In this case, the coupling
in the 2-dimensional theory would be fixed by the dimensional-reduction
process and
 would, presumably,  no longer be  minimal.
We consider this scenario next.
\subsection{Minimal Coupling in 4-D}
\par
Let us now consider a massless scalar field ($f$) that
is minimally coupled in the 4-dimensional theory. The revised
(total) action for the four dimensional theory can then be written:
\be
I^{(4)}_{TOT}=I^{(4)}-{\hbar\over 16\pi G}\int d^4x\sqrt{-g^{(4)}}
(\nabla^{(4)}f)^2,
\label{32}
\ee
where $I^{(4)}$ is the Einstein-Maxwell action of Eq.(\ref{1}). 
Again imposing spherical symmetry, we obtain a dimensionally reduced form:
\be
I_{TOT}=I-{\hbar\over 2 l^2}\int d^2 x\sqrt{-g}
\phi^2(\nabla f)^2,
\label{33}
\ee
where $I$ is the reduced action of Eq.(\ref{9}).
\par
Following the same pattern of field reparametrization and expansion as
in Section 2, we eventually find:
\be
I_{TOT}=I_{JT}-2\hbar\int d^2 x\sqrt{-{\overline g}}{\tilde \phi}
({\overline\nabla}f)^2,
\ee
where $I_{JT}$  is the JT action of Eq.(\ref{14}) (or Eq.(\ref{16}))
and we have explicitly shown the tilde and bar notation  
for the sake of clarity.
Notably, the dilaton-matter coupling  is precisely
that obtained in the dimensional reduction (from three to two dimensions) 
of a BTZ black hole, assuming minimal coupling in the higher-dimensional
theory \cite{btz,ao}. Hence, we can treat this model as a dimensionally
reduced (non-rotating) BTZ black hole.
\par
After integrating out the matter field and then taking the vacuum limit,
one can exploit the 2-dimensional trace conformal anomaly 
\cite{y0,y1,y2,y3,y4}\footnote{It has been pointed out \cite{yy1,yy2} that
some of the referenced literature contains calculational errors. However,
the demonstrated techniques  are still of considerable interest.}
 to
obtain 
 the following action \cite{on}:
\be
I_{TOT}= I_{JT} - {\hbar\over 96 \pi} \int d^2 x \sqrt{-g}
\left[ R {1\over\Box} R -{3\over\phi^2} (\nabla\phi)^2\left(
{1\over\Box}R-\ln\eta^2\right)-6\ln(\phi)R\right],
\label{666}
\ee
with tildes and bars again being suppressed.
Note that $\eta$ is an arbitrary parameter that arises out of 
the renormalization procedure \cite{bd}. 
\par
A brief aside.
It would be
remiss of us not  to point out that  such a  reduction process 
has recently been criticized.  In particular, it has been
shown that the procedures of quantization and reduction
do not necessarily commute \cite{sut}. Also,  the
reduced form of the action  may be missing non-local  terms that
originate from the conformally invariant  part of the action \cite{zg}.
However, it remains uncertain as to what (if any) effect these
considerations would have on the qualitative features of 
a first-order, semi-classical analysis.
\par
Now let us return our attentions to the non-local action of
Eq.(\ref{666}).
At this point,
one can proceed   by first re-expressing this action
in an equivalent localized form. (See, for instance,
Refs.\cite{bur2,me2}.)  This should be  followed by the careful imposition of
 boundary conditions
(which must  be suitable for a black hole in thermal equilibrium). After
which,
 the first-order stress tensor can be evaluated in a straightforward manner.
We again  defer to Refs.\cite{me1,me2} for the intricate details of
this calculation (it would also be useful to consult Refs.\cite{bur1,
fro,bur2,bur3}) and simply  quote the pertinent results:
\be
\epsilon \mu_{1}^{\prime} (x_+)= {\hbar\lambda\over 6\pi l^2},
\label{35}
\ee
\be
\epsilon \omega_{1} (x_+) = -{\hbar\over 32 \pi}\sqrt{\lambda\over l m}
\ln(4lm) + {\cal O}\left({1\over L}\right),
\label{36}
\ee
\be
\epsilon q_{1} = {\hbar l\over 96 \pi}\left[3\ln(4lm)-1\right].
\label{37}
\ee
Note that  the arbitrary (renormalization)
 constant $\eta$  has been suitably fixed  to
  eliminate it from the field equations.
\par
Substituting the above results into Eq.(\ref{30}), we obtain the
following revision to the first-order surface gravity:
\be
\kappa = \sqrt{{\lambda m\over l}} +{\hbar\lambda\over 16\pi l}
\ln(4lm)-{3\over 32} {\hbar\lambda \over \pi l}.
\label{38}
\ee
\par
We now recall the criteria that $\kappa$ be non-negative for
a legitimate black hole solution. Thus, from
the above result, it is immediately apparent that some sort
of lower bound must be imposed on the zeroth-order (but renormalized)
mass $m$.
(Notice that $\kappa\rightarrow -\infty$ as $m\rightarrow 0$.)
For illustrative purposes, let us assume that the lower
bound on mass ($m_b$) 
occurs when $4lm\sim 1$. It thus follows that:
\bea
\sqrt{m_b}&\approx& {3\hbar\over 32\pi}\sqrt{\lambda\over l}
\nonumber
\\
&\approx& {3\hbar\over 32 \pi G M_o^2}\sqrt{2M_o},
\eea
where the lower line has been expressed in terms
of the  4-dimensional parameters of $M_o$
(the mass of the black hole at theoretical extremality)
and $G$ (the Newton gravitational constant).
\par
Let us further recall that the JT black hole mass ($m$)
has been identified with $\Delta M$; that is,
the mass deviation from extremality in the RN black hole.
Hence,  the  lower bound on $m$ directly implies
that the RN black hole can {\bf not} reach
an extremal state. Furthermore, we can apply
our quantitative result for $m_b$ to determine
lower bounds on the RN  entropy and temperature
(cf. Eqs.(\ref{7},\ref{8})):
\be
S_{BH}(M,Q)\geq S_{BH}(M_o) +{3\over 8},
\label{41}
\ee
\be
T_{H}(M,Q)\geq {3\over 32\pi^2} {\hbar^2\over G^2M_o^3}.
\label{42}
\ee
\par
The situation is, of course, more complex than this on
account of the neglected logarithmic
term in Eq.(\ref{38}). (Not to mention the considerations of
Refs.\cite{sut,zg}.)  However, the qualitative  implications of
this analysis remain quite clear.
A non-extremal black hole can {\bf not} continuously
evolve to a state of extremality. Rather, the quantum back reaction
 will hinder the evaporation process;
with the black hole ultimately ``freezing'' at some
non-zero temperature that is related to the Planck
scale.  One might go so far as  to say that  back-reaction 
effects intrinsically protect the third law of thermodynamics.

\section{Conclusion}
\par

In the previous  paper, we have considered near-extremal
Reissner-Nordstrom black holes. To begin the analysis,
we reviewed a duality \cite{fnn} that exists between the near-extremal
RN sector and Jackiw-Teitelboim theory \cite{jt}. This correspondence
was established by a process of dimensional reduction, followed
by an appropriate conformal transformation of the metric
(along with a redefinition of the dilaton field that arises
out of the reduction ansatz). With a series of identifications,
it was demonstrated that the thermodynamic properties of the
JT black hole are indeed equivalent to deviations (from
extremality) in the original RN theory.
Also of note, the massless limit in  the JT solution effectively
describes the extremal limit of the RN black hole.
\par
At a first glance, the massless JT black hole appears
to be at the terminus  of a well-defined limiting procedure.
This is certainly a valid assessment
from a classical viewpoint; however,  we felt it was necessary to
clarify the situation with regard to quantum-perturbative effects.
In particular, we incorporated a (massless) quantum scalar field
and then  considered  first-order deformations to
the JT geometry.  With the assumption of  a minimally coupled
matter field in two dimensions, we found that the
well-defined massless limit remains intact. However,
the underlying assumption is decidedly incompatible
with the 4-dimensional origins of the theory. With this
in mind, we also examined the case of a minimal coupling 
 in 4-dimensions; thus giving the matter field a
4-dimensional pedigree.
\par
To begin  the revised first-order analysis, we 
considered  Einstein-Maxwell theory minimally coupled
to a quantum scalar field. After repeating the procedures
of dimensional reduction and field reparametrization,
 we found that the total action
effectively mimics that of a dimensionally reduced (non-rotating) BTZ
model \cite{btz,ao}.
With this realization, we were  able to directly apply
the results of a prior study \cite{me1,me2} and calculate
the first-order corrections to the JT geometry.
\par
Guided by a related paper \cite{aht},
we invoked a condition of non-negative surface gravity
(or, equivalently, temperature) as an appropriate litmus test
for the validity  of a low-mass JT solution.
In this manner, we were able to establish a finite lower bound on the
JT black hole mass.  For values of mass falling below this
bound, first-order perturbative effects drive the surface gravity
below zero, leading to  physically unacceptable solutions.
Given the  RN-JT duality (existing in the near-extremal RN sector),
the lower bound on JT mass directly implies  a finite lower bound
on the temperature  of a RN black hole.
That is, because of back-reaction effects, the RN black hole
will be prohibited from attaining an extremal state.
Moreover, The RN black hole can {\bf not} even come arbitrarily
close to an extremal solution; rather, it will ``freeze'' at
a finite temperature that is related to the Planck scale.
\par
Given the distinct topological differences that exist
between the extremal and non-extremal sectors \cite{hhr,tei,kr,gk},
as well as the third law of thermodynamics, it is not
surprising that a discontinuity exists between the two solutions.
What may be more of a revelation is the finite ``separation'';
that is, a non-extremal black hole
can not  come arbitrarily close to extremality.
In defense of this notion, we point out a recent investigation into
the physical spectra of charged black holes \cite{das}.  
In this study, it was generically shown that extremal
black holes can {\bf not} be achieved (at the quantum level) due to
vacuum fluctuations in the horizon. With the presumption that
such fluctuations are of Planck-scale order, this result
coincides nicely with our findings.
\par
In spite of our conclusions and similar ones elsewhere
in the literature, there remain many open questions with regard to the
status of extremal black holes. For instance, can they exist,
and (if so) can they be understood as thermodynamic systems. 
The definitive answers may have to await a  comprehensive
theory of quantum gravity.

\section{Acknowledgments}
\par
The author  would like to thank  V.P.  Frolov  for helpful
conversations. 
  \par\vspace*{20pt}

%\newpage

\end{document}